\documentclass[aps,prb,amsmath,reprint,superscriptaddress]{revtex4-1}
\usepackage{amsmath,txfonts,bm,dcolumn,graphicx,physics,color,multirow}
\usepackage[version=3]{mhchem}
\begin{document}

\title{Large-scale relativistic complete active space self-consistent field with robust convergence}
\author{Ryan D. Reynolds}
\affiliation{Department of Chemistry, Northwestern University, 2145 Sheridan Rd., Evanston, Illinois 60208, USA.}
\author{Takeshi Yanai}
\affiliation{\mbox{Institute of Transformative Bio-Molecules (WPI-ITbM), Nagoya University, Chikusa, Nagoya 464-8602, Japan}}
\affiliation{JST PRESTO, 4-1-8 Honcho, Kawaguchi-shi, Saitama 332-0012, Japan}
\author{Toru Shiozaki}
\affiliation{Department of Chemistry, Northwestern University, 2145 Sheridan Rd., Evanston, Illinois 60208, USA.}
\date{\today}

\begin{abstract}
We report an efficient algorithm using density fitting for the relativistic complete active space self-consistent field (CASSCF) method,
which is significantly more stable than the algorithm previously reported by one of the authors
[J. E. Bates and T. Shiozaki, {J. Chem. Phys.} {\bf 142}, 044112 (2015)].
Our algorithm is based on the second-order orbital update scheme with an iterative augmented Hessian procedure,
in which the density-fitted orbital Hessian is directly contracted to the trial vectors.
Using this scheme, each microiteration is made less time consuming than one Dirac--Hartree--Fock iteration, and macroiterations converge quadratically.
In addition, we show that the CASSCF calculations with the {Gaunt and full Breit} interactions can be efficiently performed by
means of approximate orbital Hessians computed with the Dirac--Coulomb {Hamiltonian}.
It is demonstrated that our algorithm can also be applied to systems under an external magnetic field, for which all of the molecular integrals are
computed using gauge-including atomic orbitals.
\end{abstract}

\maketitle

\section{Introduction}
It has long been recognized that special relativity plays an important role in chemistry, especially when heavy elements are present.\cite{Reiherbook,Liu2010MP,Saue2011CPC,Autschbach2012JCP}
The many-body relativistic quantum mechanics {can be accurately captured using the Dirac--Coulomb--Breit Hamiltonian}
\begin{align}
& \hat{H}_\mathrm{Dirac} =  \sum_i \left[c^2(\beta - I_4) + c(\boldsymbol{\alpha}\cdot\hat{\mathbf{p}}_i)
    + \hat{V}_i\right] + \sum_{i<j}\hat{V}_{ij},\\
& \hat{V}_{ij} = \frac{1}{r_{ij}} - \frac{\boldsymbol{\alpha}_{i} \cdot \boldsymbol{\alpha}_{j}}{r_{ij}}
- \frac{\left[(\boldsymbol{\alpha}_{i} \cdot \boldsymbol{\nabla}_{i})(\boldsymbol{\alpha}_{j} \cdot \boldsymbol{\nabla}_{j})r_{ij}\right]}{2},
\label{twoeterm}
\end{align}
where $\boldsymbol{\alpha}$ and $\beta$ are Dirac's matrices, {$\hat{\mathbf{p}}$ is the momentum operator, and $c$ is the speed of light.}
{The second and third terms in Eq.~\ref{twoeterm} are called the Gaunt and gauge terms, respectively; 
by neglecting either the gauge term or both we can obtain the Dirac--Coulomb--Gaunt and Dirac--Coulomb Hamiltonians.}  
In practice, since we are interested in the electronic states, the {Hamiltonian will be} projected to the electronic manifold:
\begin{align}
\hat{H}_\mathrm{ele} \approx \hat{P} \hat{H}_\mathrm{Dirac} \hat{P}.
\end{align}
There have been many studies to develop such a projector.
{
For example, the exact two-component (X2C) Hamiltonian can be obtained by a transformation that block-diagonalizes the one-electron Dirac Hamiltonian; 
the transformation is usually applied in an approximate way to the two-electron operator.\cite{Kutzelnigg2005JCP,Ilias2007JCP,Peng2013JCP}
A number of other two component approaches have also been developed.\cite{vanLenthe1993JCP,Barysz2001Theochem,Nakajima2012CR,Liu2018JCP,Seino2012JCP}
%Related to these approaches is the local transformation approach that also transforms the two-electron terms.\cite{Seino2012JCP}
The Dirac--Hartree--Fock method and
}
subsequent no-pair projection implies a projector
that block-diagonalizes the mean-field Fock operator.\cite{Visscher1994CPC,Saue1997MP,Saue1999JCP,Quiney1998AQC,Grant2000IJQC,Yanai2001JCP,Yanai2002JCP,Kelley2013JCP,Repisky2015JCTC}

Likewise, the no-pair projection based on the complete active space self-consistent field (CASSCF) method\cite{Jensen1996JCP,Thyssen2008JCP,Bates2015JCP}
decouples the Dirac equation at the CASSCF level,
which we advocate in this work{.}
%{two-component projections with the CASSCF method have also been developted.\cite{Ganyushin2013JCP,Kim2013JCP}}
The advantage of using the CASSCF-based projector is that it allows for unambiguous treatment of the two-electron operator, including the Gaunt and Breit terms if necessary.
When one performs multireference electron correlation methods for heavy-element complexes, it is natural to use this projector
because the cost of Dirac CASSCF is usually marginal compared to the multireference electron correlation treatment (such as second-order perturbation
and configuration interaction\cite{Shiozaki2015JCTC}) in terms of both operation counts and memory requirements.
We note in passing, however, that our formulation and programs reported herein are equally applicable to any relativistic Hamiltonians.

Prior to this work, a second-order four-component relativistic CASSCF algorithm without explicit construction of the entire Hessian matrix has been studied
{by Jensen and co-workers.}\cite{Jensen1996JCP,Thyssen2008JCP}
An efficient algorithm for quasi-second-order relativistic CASSCF based on density fitting has been reported by one of the authors.\cite{Kelley2013JCP,Bates2015JCP}
{Relativistic CASSCF algorithms have also been developed with a two-component projection\cite{Fleig1997TCA,Kutzelnigg2000JCP,Ganyushin2013JCP,Kim2013JCP}
or only scalar relativistic interactions included.\cite{Roos2004PCCP,Lipparini2016JCTC}}
%Another implementation of relativistic CASSCF (with spin adaptation) has recently developed by Lipparini and Gauss.\cite{Lipparini2016JCTC}
The density fitting approximation and similar techniques have also been used in non-relativistic CASSCF algorithms.\cite{Ten-no1996JCP,Aquilante2008JCP}
This work combines these developments and reports an efficient algorithm for second-order four-component relativistic
CASSCF using density fitting.
Our algorithm is optimized such that neither Hessian elements nor four-center integrals (except for those with four active indices) are constructed.
The resulting code is also applicable to molecules under a magnetic field, for which we use the gauge-including atomic orbitals\cite{London1937JPR,Wolinski1990JACS,Tellgren2008JCP,Reynolds2015PCCP} to remove the
gauge origin dependence of the results.
All of the programs have been implemented in the BAGEL package,\cite{bagel,Shiozaki2018WIREs}
which is openly available under the GNU General Public License.

\section{Theory}
We hereafter use the following orbital index notations: $i$ and $j$ label closed orbitals; $r$, $s$, $t$, and $u$ label active orbitals; $a$ and $b$ label virtual orbitals;
and $x$ and $y$ are general molecular orbitals (MOs).
$\mu$ and $\nu$ are scalar atomic orbitals.

{
The CASSCF wavefunction is parametrized using the multi-configurational \emph{ansatz}
\begin{align}
|\Psi(\mathbf{C})\rangle = \sum_I c_I |I(\mathbf{C})\rangle,
\end{align}
where each $|I(\mathbf{C})\rangle$ is a Slater determinant.  
The determination of the $c_I$-coefficients has been described elsewhere;\cite{Bates2015JCP} this work is concerned with 
the optimization of the MO coefficient matrix $\mathbf{C}$.  
}

\subsection{Augmented Hessian for complex minimax problems\label{secaug}}
In this section, the augmented Hessian approach (with scaling) for orbital updates is reviewed.
The details on the augmented Hessian algorithm for real variables have been described in depth, for instance, in Ref.~\onlinecite{Werner1987ACP}.
They have also been studied in a slightly different form in Ref.~\onlinecite{Thyssen2008JCP}.
There are extensions of this approach for non-relativistic CASSCF,\cite{Werner1981JCP,Werner1985JCP,Sun2017CPL,Ma2017JCTC}
which can similarly be translated to relativistic counterparts.

The MO coefficients are parameterized using an anti-Hermitian unitary generator $\mathbf{X}$,
\begin{align}
&\mathbf{C} = \mathbf{C}^\mathrm{ref} \exp\left( \mathbf{X} \right)\\
&X_{xy} = \left\{
\begin{array}{cll}
\displaystyle \kappa_{xy} && x > y\\
\displaystyle -\kappa^\ast_{yx} && x < y
\end{array}
\right.
\end{align}
The collection of $\kappa_{xy}$ and $\kappa_{yx}^\ast$ appearing in the expression for $\mathbf{X}$ is denoted in the following as
column vectors $\boldsymbol{\kappa}$ and $\boldsymbol{\kappa}^\ast$.
The Hessian matrix for the CASSCF energy with respect to the rotation parameters $\boldsymbol{\kappa}$ and $\boldsymbol{\kappa}^\ast$ is
\begin{align}
\mathbf{H} = \left(
\begin{array}{cc}
              \displaystyle \frac{\partial^2 E}{\partial \boldsymbol{\kappa}^\ast \partial \boldsymbol{\kappa}}
              & \displaystyle \frac{\partial^2 E}{\partial \boldsymbol{\kappa}^\ast \partial \boldsymbol{\kappa}^\ast} \\[10pt]
              \displaystyle \frac{\partial^2 E}{\partial \boldsymbol{\kappa} \partial \boldsymbol{\kappa}}
              & \displaystyle \frac{\partial^2 E}{\partial \boldsymbol{\kappa} \partial \boldsymbol{\kappa}^\ast}
\end{array}
\right) \,,
\end{align}
where the Hessian elements are evaluated at $\boldsymbol{\kappa} = \boldsymbol{\kappa}^\ast = 0$.
The Hessian matrix is complex and Hermitian.
Note that we introduced a matrix notation for the subblocks of the Hessian,
\begin{align}
\left(\frac{\partial^2 E}{\partial \boldsymbol{\kappa}^\ast \partial \boldsymbol{\kappa}}\right)_{xy,x'y'} =
\frac{\partial^2 E}{\partial {\kappa}^\ast_{xy} \partial {\kappa}_{x'y'}}.
\end{align}
{For a multi-state orbital optimization, we substitute the state-averaged CASSCF energy for $E$ throughout this derivation.}
Using a pair of trial vectors for a set of $\boldsymbol{\kappa}$ and $\boldsymbol{\kappa}^\ast$, namely $\mathbf{s}_I$ and $\mathbf{s}_J$ with
$\mathbf{s}_I = (\mathbf{t}^T_I\, \mathbf{t}^\dagger_I)^T$,
its subspace representation can be shown to be
\begin{align}
H_{IJ} \equiv \mathbf{s}_I^\dagger \mathbf{H} \mathbf{s}_J = 2\mathrm{Re}[\mathbf{t}_I^\dagger \boldsymbol{\sigma}_J],
\end{align}
where we defined the $\sigma$ vector as
\begin{align}
\boldsymbol{\sigma}_I  = \frac{\partial^2 E}{\partial \boldsymbol{\kappa}^\ast \partial \boldsymbol{\kappa}} \mathbf{t}_I + \frac{\partial^2 E}{\partial \boldsymbol{\kappa}^\ast \partial \boldsymbol{\kappa}^\ast} \mathbf{t}^\ast_I.
\label{sigmadef}
\end{align}
The right-hand side of Eq.~\eqref{sigmadef} should be understood as matrix--vector multiplications.
It is important to recognize that the subspace Hessian matrix is real and symmetric.

Given this expression, the subspace representation of the augmented Hessian matrix can be easily obtained.
The augmented Hessian is defined as
\begin{align}
\mathbf{H}' =
\left(
\begin{array}{ccc}
0 & \displaystyle \left(\frac{\partial E}{\partial \boldsymbol{\kappa}}\right)^T & \displaystyle \left(\frac{\partial E}{\partial \boldsymbol{\kappa}^\ast}\right)^T \\[10pt]
\displaystyle \frac{\partial E}{\partial \boldsymbol{\kappa}^\ast}
              & \displaystyle \frac{1}{\lambda}\frac{\partial^2 E}{\partial \boldsymbol{\kappa}^\ast \partial \boldsymbol{\kappa}}
              & \displaystyle \frac{1}{\lambda}\frac{\partial^2 E}{\partial \boldsymbol{\kappa}^\ast \partial \boldsymbol{\kappa}^\ast} \\[10pt]
\displaystyle \frac{\partial E}{\partial \boldsymbol{\kappa}}
              & \displaystyle \frac{1}{\lambda}\frac{\partial^2 E}{\partial \boldsymbol{\kappa} \partial \boldsymbol{\kappa}}
              & \displaystyle \frac{1}{\lambda}\frac{\partial^2 E}{\partial \boldsymbol{\kappa} \partial \boldsymbol{\kappa}^\ast}
\end{array}
\right)
\label{aughess}
\end{align}
where $\lambda$ is an appropriately chosen scaling factor to control the step size (see below).
We iteratively diagonalize this matrix to determine the correction vector $\Delta\boldsymbol{\kappa}$:
\begin{align}
  \mathbf{H}' \left(
  \begin{array}{c}
    1 \\ \lambda \Delta\boldsymbol{\kappa} \\ \lambda \Delta\boldsymbol{\kappa}^\ast
  \end{array}
  \right) = \varepsilon \left(
  \begin{array}{c}
    1 \\ \lambda \Delta\boldsymbol{\kappa} \\ \lambda \Delta\boldsymbol{\kappa}^\ast
  \end{array}
  \right)
\end{align}
The first trial vector is $\mathbf{s}'_0 = (1\, 0\, 0)^T$.
The other trial vectors are written as $\mathbf{s}'_I = (0\, \mathbf{t}_I^T\, \mathbf{t}^\dagger_I)^T$.
Using the above definition of $\boldsymbol{\sigma}_I$, one can show that
the subspace representation of $\mathbf{H}'$ (i.e., $H'_{IJ} = \mathbf{s}_I^{\prime\dagger} \mathbf{H}' \mathbf{s}'_J$) can be written as
\begin{align}
H'_{IJ} = \left\{
\begin{array}{cll}
0 && I = J = 0\\[10pt]
\displaystyle 2\mathrm{Re}\left[\mathbf{t}_I^\dagger \frac{\partial E}{\partial \boldsymbol{\kappa}^\ast} \right] && \mbox{$I\neq 0$ and $J = 0$}\\[10pt]
\displaystyle 2\mathrm{Re}\left[\mathbf{t}_J^\dagger \frac{\partial E}{\partial \boldsymbol{\kappa}^\ast} \right] && \mbox{$I = 0$ and $J \neq 0$}\\[10pt]
\displaystyle \frac{2}{\lambda}\mathrm{Re}\left[\mathbf{t}_I^\dagger \boldsymbol{\sigma}_J\right] && \mbox{otherwise}
\end{array}
\right.
\label{aughesssub}
\end{align}
which is a real symmetric matrix.
The lowest eigenvalue of the subspace matrix, $\varepsilon$, is related to the denominator shift in the determination of the correction vector.
The associated eigenvector elements $\{c_I\}$ define the optimal linear combination of the trial vectors, $\Delta\boldsymbol{\kappa} \approx \frac{1}{\lambda} \sum_{I\ge 1} (c_I/c_0) \mathbf{t}_I$, from which the residual vector is computed.
A new trial vector is then generated from the residual vector.
This procedure is repeated until convergence is achieved.

The value of $\lambda$ in Eq.~\eqref{aughess} is chosen such that the step size $s=|\Delta \boldsymbol{\kappa}(\lambda)|$ is maximized
within the acceptable maximum step size $s_\mathrm{max}$.
In each microiteration, we first diagonalize the augmented Hessian matrix [Eq.~\eqref{aughesssub}] and calculate the step size $s$.
If the step size is larger than $s_\mathrm{max}$, we iteratively find the optimal value for $\lambda$.
When $s$ is larger than $s_\mathrm{max}$, we set $\lambda$ to be $s/s_\mathrm{max}$;
otherwise $\lambda$ is linearly interpolated using two previous values.
When $s\le s_\mathrm{max}$ and $|s-s_\mathrm{max}| < 0.01s_\mathrm{max}$ are achieved,
we consider the value of $\lambda$ to be optimal.
As pointed out in Ref.~\onlinecite{Lipparini2016JCTC}, as long as the product $\lambda \varepsilon$ is larger than the negative eigenvalues of the Hessian associated with
the rotations between electronic and positronic orbitals (which is practically always the case), the shifted Hessian, $\mathbf{H}-\lambda \varepsilon$, has the right eigenvalue structure.
Therefore, this procedure can be applied to minimax optimization in relativistic CASSCF without problems.

The eigenvalue problem for the augmented Hessian does not have to be solved very accurately, especially when the orbital gradients are large;
we typically converge the microiteration till the root mean square (RMS) of the residual vector becomes 
{
either four orders of magnitude smaller than the step size or half of the convergence threshold specified for the macroiterations, whichever is greater.
}

\subsection{Three-index integral transformation using density fitting}
Here, we recapitulate the density fitting algorithms for four-component relativistic methods that have been reported in Ref.~\onlinecite{Kelley2013JCP}.
Density fitting approximates the scalar four-index two-electron integrals as
\begin{align}
(\mu_w \nu_w| \mu'_{w''} \nu'_{w'''}) \approx \sum_{\gamma\delta} (\mu_w \nu_{w'}|\gamma) (\mathbf{J}^{-1})_{\gamma\delta} (\delta| \mu'_{w''} \nu'_{w'''}).
\end{align}
in which $\gamma$ and $\delta$ are the auxiliary functions, and $w$ ($w=l$, $x$, $y$, and $z$) denotes the basis component,
\begin{align}
\phi_{r,w}(\mathbf{r}) =
\left\{
\begin{array}{ll}
\phi_r(\mathbf{r}) & w=l, \\[10pt]
\displaystyle{\frac{\partial \phi_r(\mathbf{r})}{\partial w}}& w = x,y,z.
\end{array}
\right.
\end{align}
The scalar integrals $(\gamma|\mu_w \nu_{w'})$ are real.
For the {Coulomb} interaction, the four-component forms of the three-index integrals can be calculated from these scalar integrals as
\begin{align}
(\gamma| \mu_X \nu_Y) = \sum_{ww'} k^{ww'}_{XY}(\gamma | \mu_w \nu_{w'}),
\end{align}
in which $X$ and $Y$ label $L^+$, $L^-$, $S^+$, and $S^-$.
The prefactor $k_{XY}^{ww'}$ is due to the use of the so-called restricted kinetic balance and is determined by
{
\begin{align}
& k_{XY}^{ww'} = \xi_X^\dagger \eta^{w\dagger} \eta^{w'} \xi_{Y}. \label{fac}
\end{align}
}
where $\xi_{X}$ is a unit vector, $\xi_{X} = (\delta_{X,L+}, \delta_{X,L-}, \delta_{X,S+}, \delta_{X,S-})^T$, and $\eta_w$ is a $4 \times 4$ matrix,
\begin{align}
&\eta^l = \begin{pmatrix}
 I_2 & 0_2 \\
 0_2 & 0_2
\end{pmatrix},
&&\eta^x = -\frac{i}{2c} \begin{pmatrix}
 0_2 & 0_2 \\
 0_2 & \sigma_x
\end{pmatrix}, \nonumber \\
&\eta^y = -\frac{i}{2c} \begin{pmatrix}
 0_2 & 0_2 \\
 0_2 & \sigma_y
\end{pmatrix},
&&\eta^z = -\frac{i}{2c} \begin{pmatrix}
 0_2 & 0_2 \\
 0_2 & \sigma_z
\end{pmatrix}. \label{Eta Matrix}
\end{align}
The details on these expressions as well as their extensions include the {Gaunt or full Breit operator} can be found in Ref.~\onlinecite{Kelley2013JCP}.

In practice, the half index transformation of the three-electron integrals is efficiently performed as
\begin{align}
(\gamma | i \nu_Y ) = \sum_X \sum_{ww'} \sum_\mu k^{ww'}_{XY} (\gamma | \mu_w \nu_{w'}) C^{X\ast}_{\mu i}.
\end{align}
where $C^X_{\nu i}$ is the $X$ block of the molecular coefficient matrix ($X=L^+$, $L^-$, $S^+$, or $S^-$).
Note that the scalar three-index integrals $(\gamma | \mu_w \nu_{w'})$ are computed at the beginning of the calculation and stored in memory;
therefore, this step consists merely of matrix--matrix multiplications and appropriate post-processing.
The second index transformation is simply
\begin{align}
(\gamma | i a ) = \sum_Y \sum_\nu (\gamma | i \nu_Y ) C^{Y}_{\nu a}.
\end{align}
These transformed integrals are extensively used in the relativistic CASSCF algorithm.
In the following, we also use the notation
\begin{align}
(\gamma_\mathbf{J} | i \nu_Y) = \sum_\delta (\mathbf{J}^{-1})_{\gamma \delta} (\delta | i \nu_Y )
\end{align}
for three index integrals that are multiplied by the metric inverse.
All of the programs for index transformation are parallelized in our code.

\subsection{Working equations using density fitting}
In this section, we explicitly note all of the working equations for contracting the Hessian matrix and a trial vector [Eq.~\eqref{sigmadef}] using the density fitting approximation.
Each subblock of the Hessian matrix can be derived by a double commutator,
\begin{subequations}
\begin{align}
&\left.\frac{\partial^2 E}{\partial \boldsymbol{\kappa}^\ast \partial \boldsymbol{\kappa}}\right|_{\boldsymbol{\kappa} = \boldsymbol{\kappa}^\ast = 0}
= \frac{1}{2} \frac{\partial^2}{\partial \boldsymbol{\kappa}^\ast \partial \boldsymbol{\kappa}} \langle 0 | [[{\hat{H}}, \hat{X}], \hat{X}] |0\rangle
\\
&\left.\frac{\partial^2 E}{\partial \boldsymbol{\kappa}^\ast \partial \boldsymbol{\kappa}^\ast}\right|_{\boldsymbol{\kappa} = \boldsymbol{\kappa}^\ast = 0}
= \frac{1}{2} \frac{\partial^2}{\partial \boldsymbol{\kappa}^\ast \partial \boldsymbol{\kappa}^\ast} \langle 0 | [[{\hat{H}}, \hat{X}], \hat{X}] |0\rangle
\end{align}
\end{subequations}
where {$\hat{H}$ is the chosen relativistic Hamiltonian} and
$\hat{X}$ is the orbital rotation generator,
\begin{align}
\hat{X}= \sum_{x> y} \left[ \kappa_{xy} \hat{E}_{xy} - \kappa^\ast_{xy} \hat{E}_{yx} \right].
\end{align}
{
To determine the sigma vector, we then contract the trial vector and its conjugate, $t_{xy}$ and $t^\ast_{xy}$, 
with the appropriate blocks of the Hessian matrix as indicated in Eq.~\ref{sigmadef}.} 

The optimized working equation is written in the matrix form so that
it maps naturally to the level-3 BLAS functions. It reads
\begin{subequations}
\label{matform}
\begin{align}
2\boldsymbol{\sigma}^{VA} &=
  2 \mathbf{f}^{VV} \mathbf{t}^{VA} \mathbf{d}^{AA}
 -\mathbf{t}^{VA} (\mathbf{d}^{AA} \mathbf{f}^{AA} + \mathbf{f}^{AA} \mathbf{d}^{AA})\nonumber\\
&-\mathbf{t}^{VC} \mathbf{F}^{CA}
+ \mathbf{F}^{VC} \mathbf{t}^{CA}
 -\mathbf{t}^{VC} \mathbf{f}^{CA} \mathbf{d}^{AA}
- 2 \mathbf{f}^{VC}\mathbf{t}^{CA} \mathbf{d}^{AA}
\nonumber\\
& - \mathbf{t}^{VA}(\mathbf{Q}^{AA}+\mathbf{Q}^{AA\dagger})
  - \mathbf{t}^{VC}\mathbf{Q}^{CA} + 2 \mathbf{R}^{VA} + 2 \mathbf{L}^{VA} \mathbf{d}^{AA} \\
2\boldsymbol{\sigma}^{CA} &=
2 \mathbf{t}^{CA} \mathbf{F}^{AA}
+ 2 \mathbf{t}^{VC\dagger}  \mathbf{F}^{VA}
- \mathbf{t}^{CA} (\mathbf{d}^{AA} \mathbf{f}^{AA} + \mathbf{f}^{AA} \mathbf{d}^{AA})
\nonumber\\
&- 2 \mathbf{F}^{CC} \mathbf{t}^{CA}
+ \mathbf{F}^{CV} \mathbf{t}^{VA}
+ 2 \mathbf{f}^{CC} \mathbf{t}^{CA} \mathbf{d}^{AA}
 -2 \mathbf{f}^{VC\dagger} \mathbf{t}^{VA} \mathbf{d}^{AA}
\nonumber\\
&- \mathbf{t}^{VC\dagger} \mathbf{f}^{VA} \mathbf{d}^{AA}
- \mathbf{t}^{VC\dagger} \mathbf{Q}^{VA}
- \mathbf{t}^{CA} (\mathbf{Q}^{AA} + \mathbf{Q}^{AA\dagger})
- 2 \mathbf{R}^{CA}
\nonumber\\
&+ 2 \mathbf{L}^{CA} - 2 \mathbf{L}^{CA} \mathbf{d}^{AA}
+ 2 \mathbf{M}^{CA} \\
2\boldsymbol{\sigma}^{VC} &= -2\mathbf{t}^{VC} \mathbf{F}^{CC}
- \mathbf{t}^{VA} \mathbf{F}^{AC}
+ 2 \mathbf{F}^{VV} \mathbf{t}^{VC}
+ 2 \mathbf{F}^{VA} \mathbf{t}^{CA\dagger}
\nonumber\\
&-  \mathbf{f}^{VA} \mathbf{d}^{AA} \mathbf{t}^{CA\dagger}
 -  \mathbf{t}^{VA} \mathbf{d}^{AA} \mathbf{f}^{AC}
 -  \mathbf{t}^{VA} \mathbf{Q}^{CA\dagger}
 -  \mathbf{Q}^{VA} \mathbf{t}^{CA\dagger} \nonumber\\
&+ 2 \mathbf{L}^{VC}
 + 2 \mathbf{M}^{VC}
\end{align}
\end{subequations}
in which the submatrices of the trial and $\sigma$ vectors are defined as
\begin{subequations}
\begin{align}
&(\mathbf{t}^{VC})_{ai} = t_{ai},
&&(\mathbf{t}^{CA})_{ir} = t^\ast_{ri},
&&(\mathbf{t}^{VA})_{ar} = t_{ar},\\
&(\boldsymbol{\sigma}^{VC})_{ai} = \sigma_{ai},
&&(\boldsymbol{\sigma}^{CA})_{ir} = \sigma^\ast_{ri},
&&(\boldsymbol{\sigma}^{VA})_{ar} = \sigma_{ar}.
\end{align}
\end{subequations}
Note the closed--active blocks are defined with their complex conjugates.
The superscripts $V$, $A$, and $C$ denote virtual, active, and closed orbitals.
The matrix notation of the density matrix within the active space is
\begin{align}
(\mathbf{d}^{AA})_{rs} = \gamma_{rs}^\ast.
\end{align}
where {use of the complex conjugate simplifies the working equations.}
{The state-averaged density matrix is substituted for multi-state calculations.} 

We also introduced the closed and generalized Fock operator ($\mathbf{f}$ and $\mathbf{F}$, respectively) in the MO basis,
\begin{subequations}
\begin{align}
&\mathbf{f} = \mathbf{h} + \mathbf{g}(\boldsymbol{\gamma}^\mathrm{closed}),\\
&\mathbf{F} = \mathbf{f} + \mathbf{g}(\boldsymbol{\gamma}^\mathrm{act}),
\end{align}
\end{subequations}
in which $\mathbf{h}$ is the one-electron Hamiltonian and the $\mathbf{g}(\boldsymbol{\gamma})$ matrix is the Coulomb and exchange part of the Fock operator associated
with the density matrix $\boldsymbol{\gamma}$,
which can be calculated using a standard Fock builder for the Dirac--Hartree--Fock method.
The submatrices of the Fock operators are also defined similarly as
$(\mathbf{f}^{CC})_{ij} = f_{ij}$, $(\mathbf{F}^{CC})_{ij} = F_{ij}$,
and so on.

The $\mathbf{Q}$ vectors in Eq.~\eqref{matform} are defined as
\begin{subequations}
\begin{align}
&(\mathbf{Q}^{CA})_{ir} = \sum_{stu} (is|tu)\Gamma_{rs,tu} \\
&(\mathbf{Q}^{AA})_{r'r} = \sum_{stu} (r's|tu)\Gamma_{rs,tu} \\
&(\mathbf{Q}^{VA})_{ar} = \sum_{stu} (as|tu)\Gamma_{rs,tu}
\end{align}
\end{subequations}
in which $\Gamma$ is the two-particle density matrix within the active space.
We further define the $\mathbf{R}$ vector, whose matrix form is
\begin{subequations}
\begin{align}
& (\mathbf{R}^{VA})_{ar} = \sum_{stu} \left[(as|t\bar{u}) + (as|\bar{t}u) + (a\bar{s}|tu)\right]\Gamma_{rs,tu} \\
& (\mathbf{R}^{CA})_{ir} = \sum_{stu} \left[(is|t\bar{u}) + (is|\bar{t}u) + (i\bar{s}|tu)\right]\Gamma_{rs,tu},
\end{align}
\end{subequations}
using the $t$-weighted coefficient matrix that defines $\bar{r}$,
\begin{align}
C_{\mu \bar{r}} = \sum_a C_{\mu a} t_{ar} - \sum_i C_{\mu i} t^\ast_{ri}.
\end{align}
All of these quantities can be efficiently computed from density-fitted molecular integrals.\cite{Bates2015JCP}

The $\mathbf{L}$ and $\mathbf{M}$ terms in Eq.~\eqref{matform} are
defined using two sets of $t$-transformed coefficients, $\bar{r}$ and $\tilde{i}$, i.e.,
$C_{\mu \bar{r}}$ above and
\begin{align}
&C_{\mu \tilde{i}} = \sum_a C_{\mu a}t_{ai} + \sum_r C_{\mu r}t_{ri}.
\end{align}
Using these $t$-transformed coefficients, we evaluate the following half-transformed integrals,
\begin{subequations}
\begin{align}
&(\gamma | \tilde{i}\nu_Y) = \sum_{X} \sum_{ww'} \sum_\mu k^{ww'}_{XY} (\gamma | \mu_w \nu_{w'}) C^{X\ast}_{\mu \tilde{i}}, \label{halftransclosed}\\
&(\gamma | \bar{r}\nu_Y) = \sum_{X} \sum_{ww'} \sum_\mu k^{ww'}_{XY} (\gamma | \mu_w \nu_{w'}) C^{X\ast}_{\mu \bar{r}}, \\
&(\gamma | \check{r}\nu_Y) = \sum_{r} (\gamma | \bar{s}\nu_Y) \gamma_{sr}.
\end{align}
\end{subequations}
The $\mathbf{L}$ and $\mathbf{M}$ matrices are computed as Fock-like contributions,
\begin{subequations}
\begin{align}
L^{XY}_{\mu\nu} &= \sum_\gamma (\gamma|\mu_X \nu_Y) \nonumber\\
&\times \left[\sum_\delta (\mathbf{J}^{-1})_{\gamma \delta}\sum_{i \mu Y'} (\delta | \tilde{i}\mu_{Y'}) C_{\mu i}^{Y'}
+ \sum_{i \mu Y'} (\gamma_\mathbf{J} | i\mu_{Y'}) C_{\mu \tilde{i}}^{Y'}\right] \nonumber\\
&- \sum_{\gamma i} \left[(\gamma_\mathbf{J}|\mu_X i) (\gamma|\tilde{i}\nu_Y) + (\gamma|\mu_X \tilde{i}) (\gamma_\mathbf{J}|i\nu_Y)\right] \label{focktclosed}\\
M^{XY}_{\mu\nu} &= \sum_\gamma (\gamma|\mu_X \nu_Y) \nonumber\\
& \times \left[\sum_\delta (\mathbf{J}^{-1})_{\gamma \delta}\sum_{r \mu Y'} (\delta | \check{r}\mu_{Y'}) C_{\mu r}^{Y'}
+ \sum_{r \mu Y'} (\gamma_\mathbf{J} | r\mu_{Y'}) C_{\mu \check{r}}^{Y'}\right] \nonumber\\
&- \sum_{\gamma r} \left[(\gamma_\mathbf{J}|\mu_X r) (\gamma|\check{r}\nu_Y) + (\gamma|\mu_X \check{r}) (\gamma_\mathbf{J}|r\nu_Y)\right]
\end{align}
\end{subequations}
These matrices are subsequently transformed to the MO basis, which yield
$\mathbf{L}^{CA}$, $\mathbf{L}^{VA}$, $\mathbf{L}^{VC}$, $\mathbf{M}^{CA}$, and $\mathbf{M}^{VC}$.

It is important to note that, in practice, the half-transformed integrals $(\gamma_\mathbf{J}|i\nu_Y)$ and $(\gamma_\mathbf{J}|a\nu_Y)$
can be computed outside the microiteration loop, because {they} do not depend on the trial vector.
The most expensive steps in each microiteration {are} then the half transformation step with the $t$-weighted closed orbitals [Eq.~\eqref{halftransclosed}]
and the Fock build using these integrals [Eq.~\eqref{focktclosed}].
The cost of the other transformations that involve active indices is almost negligible for large molecules.
Since the multiplication of the metric inverse to half-transformed integrals can be avoided, the cost of each microiteration is
typically cheaper than one Dirac--Hartree--Fock iteration in our program.

\subsection{Approximate Hessian when the {Gaunt and full Breit interactions} are used}
In second-order CASSCF algorithms, the microiteration does not have to be solved very accurately
(i.e., the Hessian elements can be slightly approximated),
as long as the convergence behavior of the macroiteration is not affected.
This is because the final CASSCF energies, which are defined as the points where the orbital gradients vanish,
remain the same even when the Hessian is approximated.
For instance, we may take advantage of this by using loose thresholds for microiterations as described above (Sec.~\ref{secaug}).
This has also been realized in the context of non-relativistic density fitted CASSCF
by Ten-no,\cite{Ten-no1996JCP} who suggested using smaller fitting basis sets for the Hessian elements used in the microiteration.

We introduce in this work an efficient scheme that is useful for
relativistic CASSCF calculations with the {Gaunt or full Breit interaction}
[i.e., calculations including the second and third terms of Eq.~\eqref{twoeterm}].
While many relativistic simulations have been performed using the standard Coulomb operator,
the Gaunt and Breit terms have to be included in some of the high-accuracy calculations
since they introduce new types of physical interactions (for instance, the direct spin--spin and spin--other-orbit interactions).
Though it has been shown by one of the authors that the Gaunt and Breit terms can be readily included in large-scale relativistic computations,\cite{Shiozaki2013JCP,Kelley2013JCP}
they do increase the computational (and memory) cost significantly, because
one has to transform the three-index integrals with large and small components.
Therefore, it is proposed that we calculate the Hessian elements using the Dirac--Coulomb {Hamiltonian}, while
the orbital gradients are calculated { with the Gaunt or Breit terms included;}
in other words, we replace Eq.~\eqref{aughess} with
\begin{align}
\mathbf{H}' =
\left(
\begin{array}{ccc}
0 & \displaystyle \left(\frac{\partial E}{\partial \boldsymbol{\kappa}}\right)^T & \displaystyle \left(\frac{\partial E}{\partial \boldsymbol{\kappa}^\ast}\right)^T \\[10pt]
\displaystyle \frac{\partial E}{\partial \boldsymbol{\kappa}^\ast}
              & \displaystyle \frac{1}{\lambda}\frac{\partial^2 E_\mathrm{DC}}{\partial \boldsymbol{\kappa}^\ast \partial \boldsymbol{\kappa}}
              & \displaystyle \frac{1}{\lambda}\frac{\partial^2 E_\mathrm{DC}}{\partial \boldsymbol{\kappa}^\ast \partial \boldsymbol{\kappa}^\ast} \\[10pt]
\displaystyle \frac{\partial E}{\partial \boldsymbol{\kappa}}
              & \displaystyle \frac{1}{\lambda}\frac{\partial^2 E_\mathrm{DC}}{\partial \boldsymbol{\kappa} \partial \boldsymbol{\kappa}}
              & \displaystyle \frac{1}{\lambda}\frac{\partial^2 E_\mathrm{DC}}{\partial \boldsymbol{\kappa} \partial \boldsymbol{\kappa}^\ast}
\end{array}
\right)
\end{align}
in which $E_\mathrm{DC}$ is the CASSCF energy expression with the Dirac--Coulomb interaction.
This procedure allows us to reduce the cost of microiterations considerably.
It will be shown later that this procedure does not deteriorate the quadratic convergence behavior.

\subsection{Relativistic CASSCF with a magnetic field}
When molecules are placed under an external magnetic field, the interaction between molecules and the magnetic field
can be described by replacing the momentum operator $\mathbf{p}$ with {$\boldsymbol{\pi} = \mathbf{p} + \mathbf{A}(\mathbf{r})$} where
$\mathbf{A}(\mathbf{r})$ is the vector potential generated by the external magnetic field; for a uniform field $\mathbf{B}$ in the Coulomb gauge,
\begin{align}
\mathbf{A}(\mathbf{r}) = \frac{1}{2} \mathbf{B} \times (\mathbf{r}-\mathbf{r}_G),
\end{align}
in which $\mathbf{r}_G$ is the gauge origin.
The interaction with a magnetic field breaks the time-reversal symmetry.
Furthermore, to remove the gauge origin dependence of the results, one has to use the so-called gauge-including
atomic orbitals (GIAOs) as a basis:
\begin{align}
\phi^\mathrm{GIAO}_x (\mathbf{r}) = \phi_x (\mathbf{r}) \exp\left(-i \mathbf{A}(\mathbf{R})\cdot \mathbf{r} \right),
\end{align}
in which $\mathbf{R}$ is the position of the basis function.
The small component basis functions are constructed using the
{restricted magnetic balance relation,\cite{Aucar1999JCP,Kutzelnigg2003PRA,Komorovsky2008JCP} rather than the standard restricted kinetic balance.  }
The evaluation of molecular integrals in relativistic calculations using the GIAOs have been reported by Reynolds and Shiozaki,\cite{Reynolds2015PCCP}
in which the computational cost of performing Dirac--Hartree--Fock calculations with GIAOs has been shown to be almost identical to
the one with the standard basis sets, because the equations in the relativistic framework (even in the absence of magnetic fields) are complex.

Here, we extended the relativistic CASSCF program described above so that it can handle molecules
placed under a magnetic field.
Since we do not utilize the time-reversal symmetry in the active FCI code and
in the integral transformation programs, this extension has been trivial;
the only difference is that we disable the time-reversal adaptation of the molecular coefficient matrix
when molecular coefficients are updated in the macroiteration.
It will be shown in the section below that the rapid convergence is observed when starting from
the orbitals obtained from the calculation without a magnetic field.

\section{Results}

\begin{figure}[t]
\includegraphics[width=0.48\textwidth]{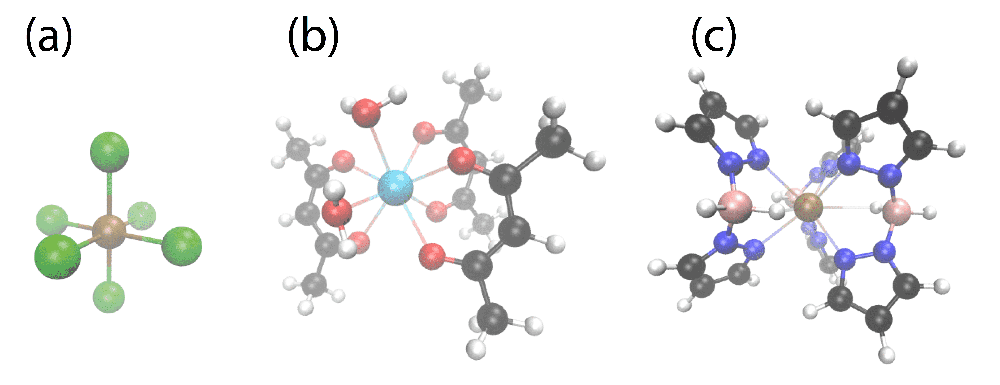}
\caption{Geometries of molecules used for benchmarking.
    (a) \ce{UCl6}
    (b) \ce{Dy(Acac)3(H2O)2}
    (c) \ce{U(H2BPz2)3}
} \label{structures}
\end{figure}

\begin{figure*}[t]
\includegraphics[width=0.80\textwidth]{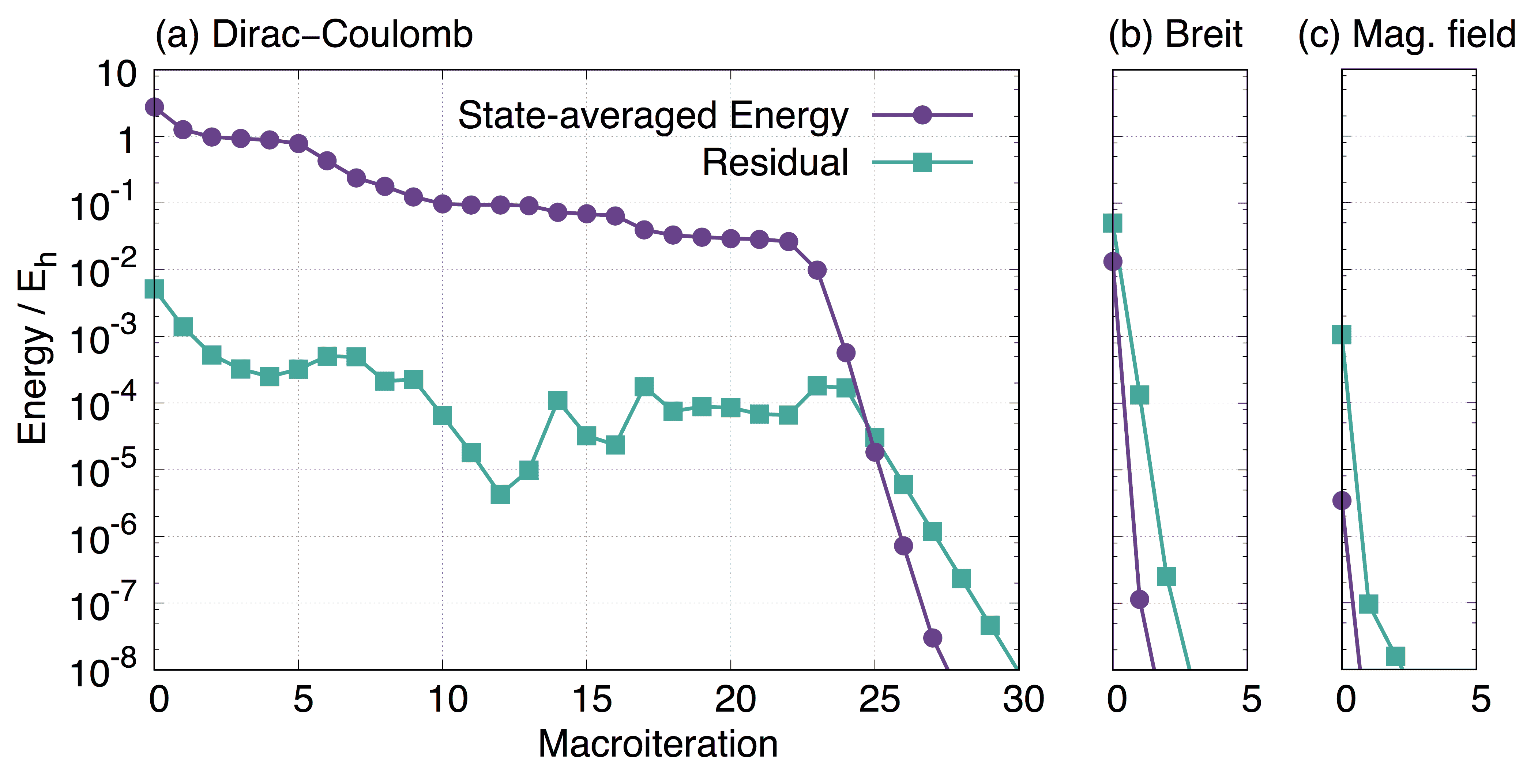}
\caption{Convergence of \ce{Dy(Acac)3(H2O)2} energy and residual with each Dirac CASSCF macroiteration.
    (a) Dirac--Coulomb Hamiltonian with zero magnetic field, starting from loosely converged Dirac--Hartree--Fock orbitals.
    (b) {Dirac--Coulomb--Breit} Hamiltonian with zero magnetic field, starting from the converged orbitals of (a).
    (c) Dirac--Coulomb {Hamiltonian} with a 20 T axial magnetic field, starting from the converged orbitals of (a).
} \label{convergence-dyacac}
\end{figure*}

\subsection{Convergence Benchmarks} \label{convergence}

To show the robustness of our algorithm, we present the convergence behavior for a challenging system, the 49-atom Dysprosium acetylacetone complex,
\ce{Dy(Acac)3(H2O)2}, shown in Figure~\ref{structures}(b).
The calculations used an active space of 9 electrons in the 7 $f$-orbitals.
For the Dirac--Coulomb CASSCF calculation, initial guess orbitals were obtained from a closed-shell Dirac--Hartree--Fock calculation performed on
the +9 cation and converged to a residual norm of $10^{-5}$ a.u.
The energy minimized in the orbital optimization was the average of the 16 states in the $J=15/2$ multiplet.
{
A mixed ANO-RCC basis set\cite{Widmark1990TCA,Roos2004JPCA,Roos2008JPCA} was used for these benchmarks:  
the uncontracted basis set was applied to the dysprosium atom, the TZP contraction to all oxygen atoms, and the DZP contraction to the carbon and hydrogen atoms.  
}
The fitting basis set used for light atoms was uncontracted TZVPP-JKFIT for light atoms\cite{Weigend2008JCC} when the Dirac--Coulomb Hamiltonian is used.
For Breit calculations, one additional $p$, $d$, and $f$ function was added to the auxiliary basis set for C and O by
multiplying the exponent of the previous tightest function by 2.5.
The fitting basis set for Dy was generated by decontracting ANO-RCC, generating an $i$-shell by copying the exponents
from the existing $h$-shell, and then adding 4$s$, 4$p$, 7$d$, 8$f$, 6$g$, 6$h$, and 5$i$ functions, each tighter than the last by a factor of 2.5.
The auxiliary basis for H and Dy was not changed between Coulomb and Breit Hamiltonians.
All calculations used a Gaussian distribution to represent the finite nuclear charge,{\cite{Visscher1997ADNDT}}
 and either restricted kinetic or restricted magnetic
balance for the small component basis functions, as appropriate.  All calculations with magnetic field used a GIAO basis.
The convergence behavior was tested using tight convergence thresholds:  the residual norm was converged to $10^{-10}$ for the active CI part,
and the total convergence threshold was set to $10^{-8}$.
The microiterations were converged to a residual norm of $10^{-4}$ multiplied by the step size, except for part (c) where this factor was
loosened to $5\times10^{-3}$.

The convergence behavior of the new CASSCF algorithm for {the Dysprosium acetylacetone complex} is illustrated in Figure~\ref{convergence-dyacac}.
The relative state-averaged energy and residual norm are plotted against the number of macroiterations elapsed.
Note that the initial CASSCF energy was higher by many Hartrees than the converged energy {due to} the use of poor initial guess orbitals.
Even for this challenging case, our program smoothly lowers the energy towards the solution; once it reaches the quadratic region (at around 22nd iteration),
we observe the second-order convergence as expected.
Figure~\ref{convergence-dyacac}(b) and (c) show the convergence behavior when either the Breit {operator} or a 20 Tesla axial magnetic field is added,
using the converged orbitals from part (a) as the starting guess.
Both of these results show very rapid convergence to the solutions, since the initial guess orbitals from the Dirac--Coulomb calculation
without a magnetic field were within the quadratic region in {each} case.

\begin{table*}[tb]
\caption{Average {wall} times, in seconds, for the molecules tested in this work.
All timing benchmarks used two MPI processes per 16-core compute node, with 8 threads per process.
\label{timing}
}
\begin{ruledtabular}
\begin{tabular}{llcrrrdddd}
 Molecule & Hamiltonian & Active & Closed & Virtual & Nodes\footnotemark[1] & \multicolumn{2}{c}{Micro\footnotemark[2]} & \multicolumn{1}{c}{Active FCI} & \multicolumn{1}{c}{Macro} \\ \hline
{\ce{UCl6}}
       & Coulomb                            & (6,6)   &  94 &  453 &  4 &  83.8 &(11.8) &  96.7 & 1133.1 \\      % NMacro = 17
       & Coulomb                            & (12,12) &  91 &  450 &  4 &  89.2 &(34.1) & 508.3 & 3613.5 \\      % NMacro = 20
       & Coulomb                            & (6,6)   &  94 &  993 &  8 & 165.7 &(12.5) & 142.1 & 2289.4 \\[5pt] % NMacro =  6
{\ce{Dy(Acac)3(H2O)2}}
       & Coulomb                            & (9,7)   & 118 &  751 & 24 & 100.6 &(17.7) & 113.2 & 1984.8 \\      % NMacro = 20
       & { Coulomb + Gaunt }    & (9,7)   & 118 &  751 & 24 & 111.8 &(17.3) & 316.5 & 2537.9 \\      % NMacro =  3
       & Coulomb + Breit                    & (9,7)   & 118 &  751 & 24 & 122.6 &(17.3) & 831.0 & 3359.0 \\      % NMacro =  3
       & Coulomb + Magnetic                 & (9,7)   & 118 &  751 & 24 & 124.4 &(15.0) & 165.3 & 2123.2 \\[5pt] % NMacro =  1
{\ce{U(H2BPz2)3}}
       & Coulomb                            & (3,7)   & 160 &  764 & 48 & 122.0 &(18.1)  & 162.1& 2465.6 \\      % NMacro = 13
       & { Coulomb + Gaunt }    & (3,7)   & 160 &  764 & 48 & 128.4 &(15.3)  & 389.9& 2694.0 \\      % NMacro =  3
       & Coulomb + Breit                    & (3,7)   & 160 &  764 & 48 & 143.6 &(16.0)  &1129.0& 3912.2 \\      % NMacro =  3
\end{tabular}
\end{ruledtabular}
\footnotetext[1]{Each node consists of two Xeon E5-2650 2.00GHz CPUs (Sandy Bridge, 16 CPU cores per node; purchased in 2012) with InfiniBand QDR.}
\footnotetext[2]{The numbers in parentheses are the average number of microiterations per macro iteration.}
\end{table*}

\subsection{Timing Benchmarks}

Timing benchmarks have been performed on {several} molecules containing heavy elements.  
The structures of the three molecules used are shown in Figure~\ref{structures}.  
The smallest molecule tested is the octahedral complex \ce{UCl6} with U-Cl bond lengths of 2.42~{\AA} for the singlet ground state.\cite{Zachariasen1948AC}
The point group symmetry was not used in the calculations.   
Two larger organometallic single-molecule magnets, \ce{Dy(Acac)3(H2O)2} and \ce{U(H2BPz2)3}, were also included in the timing benchmarks.
Both of these single-molecule magnets were tested at their experimental structures,\cite{Jiang2010ACIE,Rinehart2010JACS}
with hydrogen atom positions {optimized} using density functional theory (DFT)
with the PBE functional.\cite{Perdew1996PRL}
The DFT optimization was performed by the Turbomole package,\cite{Ahlrichs1989CPL,Furche2014WIREs}
{
and used the def2-TZVPP basis set except for Dy and U atoms, which used def-TZVPP with effective core potential.\cite{Weigend1998CPL,Dolg1989JCP,Kuechle1994JCP,Cao2003JCP} 
Resolution of the identity DFT (RI-DFT) was used with the corresponding RI-J auxiliary basis sets.\cite{Eichkorn1997TCA,Weigend2006PCCP}
}
The resulting molecular coordinates are included in the supplementary material.
The calculations performed here are for the lowest 16 states in \ce{Dy(Acac)3(H2O)2} and lowest 10 states in \ce{U(H2BPz2)}, which 
correspond to the ground $J=15/2$ and $J=9/2$ multiplets.  

In each case, CASSCF calculations were first started from loosely converged Dirac--Hartree--Fock orbitals.  
Subsequent calculations used these converged orbitals as the starting guess and added another factor, be it the {Gaunt or} Breit 
interaction, an external magnetic field, or a larger basis set or active space.  
The orbital basis sets for each calculation were uncontracted ANO-RCC basis sets for the central $f$-block atom and contracted 
ANO-RCC basis sets for the ligands; 
{
the TZP contraction was used for oxygen and chlorine atoms, 
and the DZP contraction was used for hydrogen, carbon, boron, and nitrogen.\cite{Roos2008JPCA,Roos2005CPL,Widmark1990TCA,Roos2004JPCA}
}
The auxiliary basis sets for density fitting were the same as described in the previous section: uncontracted TZVPP-JKFIT where possible 
and {the} large customized fitting basis set {for $f$-block elements.}  Whenever the {Gaunt or} Breit interaction was included, the auxiliary basis 
for C, B, N, O, and Cl atoms was supplemented by adding one tighter $p$, $d$, and $f$ function with an exponent 2.5 times higher than the previous tightest 
function in the angular shell.  
An overall convergence threshold for the residual norm of 10$^{-7}$ was used, while the 
convergence threshold for active FCI was set to 10$^{-10}$.  
As in the convergence benchmarks in Sec.~\ref{convergence}, the convergence threshold for microiterations was normally set to 
$10^{-4}$ multiplied by the step size, but this was loosened to $5\times10^{-3}$ when a magnetic field was applied.  
For calculations with the Breit operator, we reduced the memory cost by splitting up the contributions of 
closed orbitals to the Fock operator into smaller batches.  (The default in BAGEL is to treat up to 250 spin-orbitals simultaneously; 
this parameter was reduced to 150 for the Breit calculations.) 

Each calculation was run to convergence, and the average timings for each macroiteration are shown in Table~\ref{timing}.  
The wall times required for the average microiteration, macroiteration, and active FCI part are shown for each calculation.  
The total cost of a macroiteration is slightly greater than the sum of these contributions.
The active FCI timing includes calculation of the closed Fock matrix with updated orbitals, iterative solution for CI amplitudes, and 
computation of reduced density matrices.  The total cost is generally dominated by the microiterations, with the active FCI part accounting for 
at least 60\% of the remaining cost, followed by smaller costs associated with the integral transformations and 
computation of the $\mathbf{Q}$ vectors and orbitals gradients.  
The number of microiterations needed to obtain the Hessian varies substantially between calculations and sometimes from one macroiteration to the next, 
while the other costs per iteration vary only slightly throughout a calculation.  

For \ce{UCl6}, we tested two active spaces.  The smaller active space is that consisting of six electrons in six orbitals, which correspond to
$t_{1u}$ $\sigma$-bonding and antibonding orbitals {(although symmetry is not enforced)}.  
The larger active space adds six more to {incorporate} $\pi$-donation into the $f$-orbitals.  
It was observed that the increase in the size of the active spaces raised the per-iteration timing by more than a factor of three, 
largely due to slower convergence in the iterative Hessian solver.  
We also measured the timing for the smaller active space with a large basis set {generated} by
decontracting ligand basis orbitals.
The use of the large basis set increased the cost of each step, but it did not lead to the same increase in the
number of microiterations.  

Using the two larger molecules, we assessed the cost associated with the use of the {Gaunt and Breit interactions}.  
The computational cost associated with each {micro}iteration increased by 
{5-12\% for Dirac--Coulomb--Gaunt and approximately 20\% for Dirac--Coulomb--Breit.
The higher cost of the Breit operator is partly due to the sub-optimal
division} of closed orbitals, which was necessary to reduce the memory cost so that the calculations could be run using the same number of compute nodes.  
The cost of the active FCI part increased substantially, by { a factor of 2-3 for Gaunt and about 7 for Breit}, 
largely due to the expense of computing the MO Hamiltonians in the active space. 
When a magnetic field was applied to \ce{Dy(Acac)3(H2O)2}, we observed only a small increase in cost owing to the use of complex atomic orbital basis functions.  
The numbers of microiterations for these test cases are reported to be smaller than {for} the Dirac--Coulomb calculations, but this is due to the fact that
the starting orbitals (which are the solutions of the Dirac--Coulomb calculations) are of higher quality.

\section{Conclusions}
We have developed an efficient algorithm for four-component relativistic CASSCF, which is applicable to sizable systems and
is significantly more stable than the previously reported density-fitted algorithm.\cite{Bates2015JCP}
The new algorithm is based on the augmented Hessian updates using density fitting. Each microiteration has been made slightly cheaper than one Dirac--Hartree--Fock iteration.
The working equation is reported in Eq.~\eqref{matform}, which is the main result of this work.
We have also introduced a scheme to speed up the relativistic CASSCF calculations with the {Gaunt and full Breit terms included in the Hamiltonian},
in which the Hessian matrix elements are computed with the Dirac--Coulomb operator.
Furthermore, we have shown that the code can be applied to systems under a magnetic field, in which gauge-including atomic orbitals are used.
All of the programs are parallelized using both threads and MPI processes
and are part of the BAGEL program package distributed under the GNU General Public License.

{
\section{Supplementary Material}

See supplementary material for the coordinates of the two large molecules used for timing analysis, as well as for the converged total energies 
obtained from the timing benchmarks.   
}

\section{Acknowledgments}
RDR has been supported by the DOD National Defense Science and Engineering Graduate (NDSEG) Fellowship, 32 CFR 168a.
TS has been supported by the National Science Foundation CAREER Award (CHE-1351598).
The development of the program infrastructure in BAGEL has been in part supported by National Science Foundation (ACI-1550481).
TY has been supported by JSPS KAKENHI (JP16H04101 and JP15H01097) and JST PRESTO (JP17937609).

\end{document}